\documentclass[aps,11pt,nofootinbib,showpacs,showkeys]{revtex4}
\usepackage{amsmath}
\usepackage{amsfonts}
\usepackage{amssymb}
\usepackage{graphicx}%
\begin{document}

\title{Cosmological analytic solutions with reduced relativistic gas}
\author{L. G. Medeiros}\thanks{leogmedeiros@ect.ufrn.br}
\affiliation{Escola de Ci\^{e}ncia e Tecnologia, Universidade
Federal do Rio Grande do Norte.} \affiliation{Campus Universitário
s/n, CEP 59072-970, Natal, Brazil}

\begin{abstract}
In this paper one examine analytical solutions for flat and non-flat universes
composed by four components namely hot matter (ultra-relativistic), warm
matter (relativistic), cold matter (non-relativistic) and cosmological
constant. The warm matter is treated as a reduced relativistic gas and the
other three components are treated in the usual way. The solutions achieved
contains one, two or three components of which one component is of warm matter
type. A solution involving all the four components was not found.

\end{abstract}


\maketitle

\section{Introduction}

At recent years cosmology has lived a gold age. Many observational techniques
are being developed and they are producing a lot of data about our universe.
We can cite for example techniques involving observation of supernovas
\cite{SN1,SN1A,SN2,SN2A,SN3}, detection of cosmic microwave background anisotropies
\cite{CMB1} and surveys of galaxies \cite{Sur1,Sur2,Sur3}. These new data have
generated a picture where the universe is constituted by five components:
radiation, neutrinos, baryonic matter, dark matter and dark energy. In the
simplest models the components evolve independently which means each one has
its one equation of state.

In the standard approach the evolution of the universe is divided in
eras, each one dominated by one component. Separate solutions are
used for each era and intermediate periods are connected matching
the initial conditions. This approach, although approximated,
describes the history of the universe in a simple way. However, the
precision of new data, make it desirable to have solutions as
complete as possible.

Analytic solutions involving cold matter, radiation and cosmological constant
were studied in \cite{Solutions,Ref1,Ref2,Ref3}. Nevertheless, none of these
articles consider relativistic components in their analysis. In principle, if
we look for solutions that take into account relativistic particles it would
be necessary to deal with equations of state containing modified Bessel
functions. This would make it very difficult to obtain analytical solutions.
Fortunately, in 2005 it was proposed a simpler formulation to describe
relativistic particles \cite{RRG1}. This formulation, known as reduced
relativistic gas (RRG), is able to represent a gas of relativistic particles
with good accuracy. Besides, the RRG model is simple enough which allows us to
search for analytic solutions to Friedmann equations.

The first analytic solutions containing the RRG were found in
\cite{RRG1}. Here one continue this work extending the analysis for
cases involving RRG with other components. It is discussed solutions
for a universe composed by a RRG component plus non-relativistic
matter, radiation and/or cosmological constant. These kind of
solutions are important whenever we want to describe a universe
which has a component with relativistic behavior. Good examples are
models which involves warm dark matter \cite{RRG2,RRG3}.

The paper is organized as follows. In section $2$ is given a summary
about RRG focusing in its connection with standard cosmology. The
content of this section is a resume of sections $2$ and $3$ of
\cite{RRG1}. In section $3$ is presented analytic solutions for a
universe containing one, two and three components where one of these
components is modeled by RRG. In general, the flat and non-flat
cases were treated separately. The final comments and further
perspectives are given in section $4$.

\section{Reduced Relativistic Gas in the Standard Cosmology}

The RRG is a simple model for a relativistic ideal gas of massive
particles and it was first introduced in \cite{RRG1}. The idea
behind of this model is to use the kinetic theory of gases attached
with relativistic concepts. Using standard considerations which
relate the transferred relativistic moment by particles with the
pressure $p$ produced in a wall, allow us to write
\[
p=\frac{1}{3V}\frac{mv^{2}}{\sqrt{1-v^{2}/c^{2}}},
\]
where $V$ is the volume, $m$ is the mass and $v$ is the relativistic
velocity.

Supposing that all particles have the same relativistic kinetic energy
$\varepsilon$ one can rewrite the equation above as%
\begin{equation}
p=\frac{\rho}{3}\left[  1-\left(  \frac{nmc^{2}}{\rho}\right)  ^{2}\right]
=\frac{\rho}{3}\left[  1-\left(  \frac{\rho_{d}}{\rho}\right)  ^{2}\right]
\label{EoS_1}%
\end{equation}
where $\rho=n\varepsilon$ is the energy density and $n$ is the numerical
density of particles. Note that $\rho_{d}$ is the energy density of
non-relativistic particles and thus it is proportional to $V$, i.e. $\rho
_{d}=\rho_{1}V$ . It is easy to see that if $\rho\simeq\rho_{d}$
(non-relativistic particles) the equation of state (\ref{EoS_1}) reduces to
$p\simeq0$, and if $\rho\gg\rho_{d}$ (ultrarelativistic particles) the
equation of state (EoS) becomes $p=\rho/3$.

It is instructive compare (\ref{EoS_1}) with the correct EoS derived by the
statistical mechanics of ensembles. Computing the partition function of
classical relativistic ideal gas we can determine $p$ and $\rho$ as functions
of $n$ and $kT$:%
\begin{align*}
p  &  =nkT\\
\rho &  =nmc^{2}\frac{K_{3}\left(  mc^{2}/kT\right)  }{K_{2}\left(
mc^{2}/kT\right)  }-nkT
\end{align*}
where $K_{v}$ is a modified Bessel function of index $\nu$. Combining this two
equations we obtain%
\begin{equation}
\rho=\rho_{d}\frac{K_{3}\left(  \rho_{d}/p\right)  }{K_{2}\left(  \rho
_{d}/p\right)  }-p. \label{EoS_2}%
\end{equation}

At first sight, (\ref{EoS_1}) and (\ref{EoS_2}) are completely different.
However, a numerical comparison between these two equations was performed in
\cite{RRG1} and there it was shown they are quite similar. Indeed, the
difference between (\ref{EoS_1}) and (\ref{EoS_2}) is at most $2.5\%$ and
becomes negligible at ultrarelativistic and non-relativistic regimes. Thus,
(\ref{EoS_1}) is a good approximation for the EoS of classical relativistic
ideal gas with the great benefit of being much simpler than (\ref{EoS_2}).

In order to use (\ref{EoS_1}) as an EoS for a relativistic component of cosmic
fluid it is necessary determined how the energy density depends on scale
factor. This is performed writing the covariant conservation law in terms of
volume%
\begin{equation}
\frac{d\rho}{dV}=-\frac{\left(  \rho+p\right)  }{V}\text{ \ \ \ \ \ where
\ \ \ \ }V\sim a^{3}, \label{Cons law}%
\end{equation}
and replacing (\ref{EoS_1}) in (\ref{Cons law}). Solving the
differential equation we obtain
\begin{equation}
\rho_{RRG}\left(  a\right)  =\sqrt{\rho_{1}^{2}\left(  \frac{a_{0}}{a}\right)
^{6}+\rho_{2}^{2}\left(  \frac{a_{0}}{a}\right)  ^{8}}. \label{rho_a}%
\end{equation}
where the initial condition used was $\rho_{RRG}\left( a_{0}\right)
=\sqrt{\rho_{1}^{2}+\rho_{2}^{2}}$.

\begin{figure}[th]
\begin{center}
\includegraphics[width=13cm]{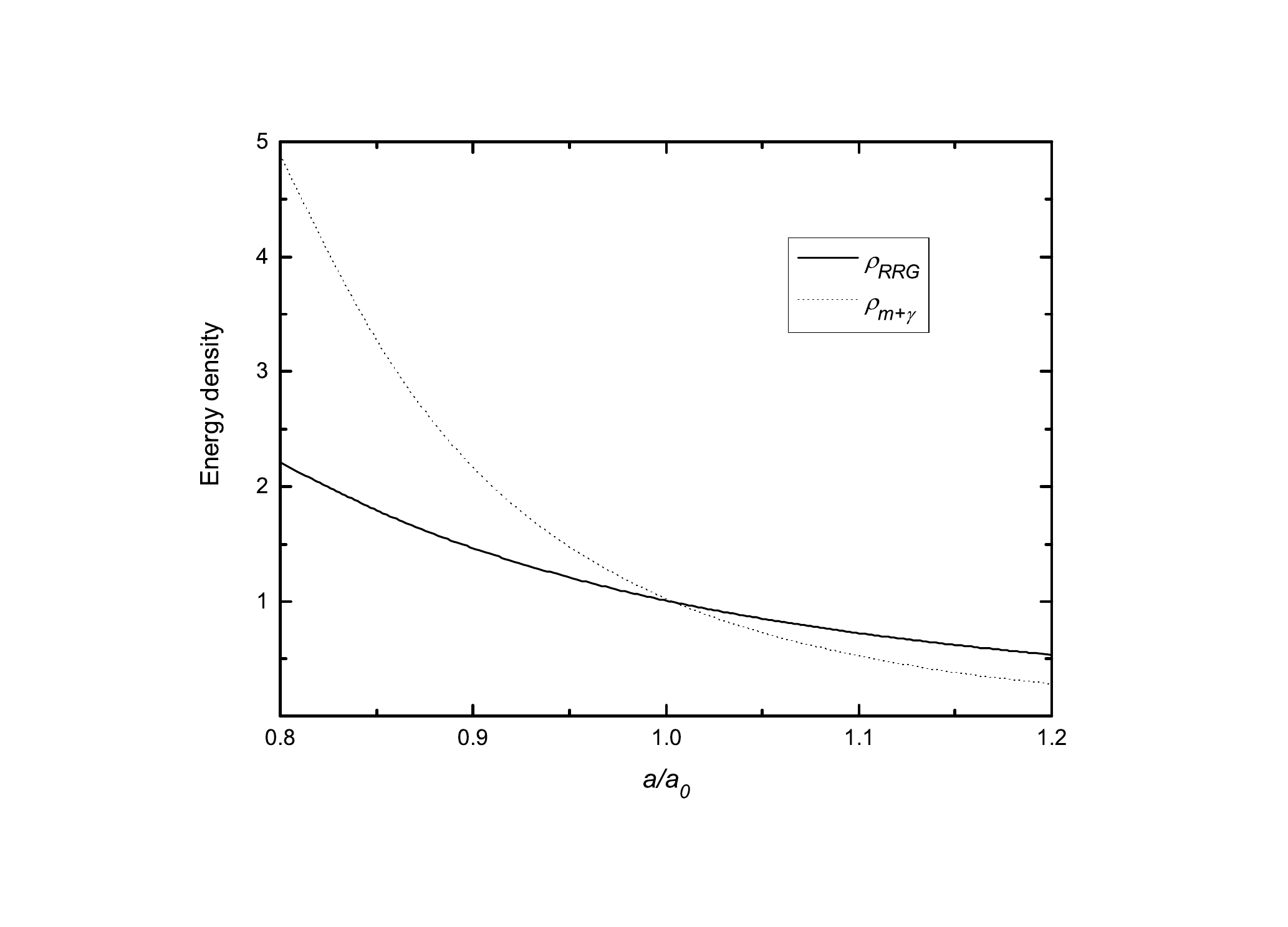}
\end{center}
\caption{The energy density of $\rho_{RRG}$ (full line) and
$\rho_{m+\gamma}$ (dashed line) in terms of $a/a_{0}$. In both case it was chosen $\rho_{1}%
=\rho_{2}$ with the initial condition $\rho\left(  a_{0}\right)
=1$. Observe that the transition among the radiation and dust behavior is smoother for $\rho_{RRG}$.}%
\label{fig1}%
\end{figure}

Analyzing the last equation we can associate the constants $\rho_{1}$ and
$\rho_{2}$ as the energy densities of dust and radiation respectively. Indeed,
if we take $\rho_{2}=0,$ (\ref{rho_a}) scaling as a dust-like component which
means $\rho\sim a^{-3}$. And if we take $\rho_{1}=0,$ (\ref{rho_a}) scaling as
a radiation-like component which means $\rho\sim a^{-4}$. Although,
(\ref{rho_a}) reproduces these two behaviors it is qualitatively and
quantitatively different from a cosmic fluid composed by dust and radiation.
In the first case, we have a single relativistic component represented by
(\ref{rho_a}), and in the second case, we have two distinct components whose
the energy density is given by%
\begin{equation}
\rho_{m+\gamma}\left(  a\right)  =\rho_{1}\left(  \frac{a_{0}}{a}\right)
^{3}+\rho_{2}\left(  \frac{a_{0}}{a}\right)  ^{4}. \label{m_gama}%
\end{equation}
A numerical confrontation between (\ref{rho_a}) and (\ref{m_gama}) is shown in
figure \ref{fig1}.%


\section{Analytic Solutions}

Suppose that the cosmic fluid are composed by four independent components
namely radiation ($\gamma$), cold matter ($CM$), cosmological constant
($\Lambda$) and warm matter ($WM$). Thus, the first Friedmann equation in
units of $c=1$ results in
\begin{equation}
\left(  \frac{\dot{a}}{a}\right)  ^{2}=\frac{8\pi G}{3}\left[  \rho_{\gamma
0}\left(  \frac{a_{0}}{a}\right)  ^{4}+\rho_{CM0}\left(  \frac{a_{0}}%
{a}\right)  ^{3}+\frac{\Lambda}{8\pi G}+\sqrt{\rho_{1}^{2}\left(  \frac{a_{0}%
}{a}\right)  ^{6}+\rho_{2}^{2}\left(  \frac{a_{0}}{a}\right)  ^{8}}\right]
-\frac{\kappa}{a^{2}}, \label{Fried_1}%
\end{equation}
where $\kappa$ is the spatial curvature and the $WM$ is modeled by the RRG.
Usually, the component of $WM$ represents a warm dark matter (as proposed in
\cite{RRG2,RRG3}). Nevertheless, another physical possibility is use the $WM$
component to describe neutrinos.

The main goal of this work is to study the analytical solutions linked with
(\ref{Fried_1}). To perform this it is convenient to define the following
quantities:%
\begin{align}
\Omega_{\gamma0}  &  =\frac{\rho_{\gamma0}}{\rho_{c}}\text{, \ }\Omega
_{CM0}=\frac{\rho_{CM0}}{\rho_{c}}\text{, \ }\Omega_{\Lambda}=\frac{\Lambda
}{3H_{0}^{2}}\text{,}\nonumber\\
\ \Omega_{1}  &  =\frac{\rho_{1}}{\rho_{c}}\text{, \ }b=\frac{\rho_{2}}%
{\rho_{1}}\text{ and \ }\Omega_{\kappa0}=\frac{\kappa}{H_{0}^{2}}\text{ \ with
\ }\rho_{c}=\frac{3H_{0}^{2}}{8\pi G}. \label{def Omegas}%
\end{align}
Since (\ref{Fried_1}) is a separable differential equation of first order it
can be written as an integral in the scale factor. Thus, using
(\ref{def Omegas}) we obtain,%
\begin{equation}
t\left(  a\right)  =\pm\frac{1}{H_{0}} {\displaystyle\int}
\frac{ada}{\sqrt{\Omega_{\gamma0}+\Omega_{CM0}a+\Omega_{\Lambda}a^{4}%
+\Omega_{1}\left[  a^{2}+b^{2}\right]  ^{1/2}-\Omega_{\kappa0}a^{2}}},
\label{Fried_Int}%
\end{equation}
where $a_{0}\equiv1$. The sign will be chosen so to get always an expanding universe.

The approach adopted here is to search solutions involving one, two and three
components with and without curvature\footnote{An analytic solution involving
all the four components was not achieved.}. The integral in (\ref{Fried_Int})
can be written as a sum of integrals whose its structures are of type
\[
{\displaystyle\int} F\left(  a,\sqrt{P\left(  a\right)  }\right)
\]
where $P\left(  a\right)  $ is a polynomial. Associated with this integral we
have three possibilities \cite{Grad}:

\begin{enumerate}
\item[(i)] If $P\left(  a\right)  $ is at most second-degree polynomial and
$F$ has a simple structure then the solution of (\ref{Fried_Int}) can be
expressed in terms of algebraic functions. In this case, an explicit expression
sometimes can be found.

\item[(ii)] If $P\left(  a\right)  $ is third- or fourth-degree polynomial
and/or $F$ has not a simple structure then the solution of (\ref{Fried_Int})
can be written in terms of elliptical integrals (see appendix \ref{Ap}). In
this case, the solution is only implicit.

\item[(iii)] If $P\left(  a\right)  $ is more than fourth-degree polynomial
then it is not possible to obtain a solution for (\ref{Fried_Int}) - e.g. the
integral with all four components.
\end{enumerate}

Before to move on for the specific cases it is noteworthy that the solutions
without $WM$ is not treated in this paper. This kind of solution was
extensively studied in \cite{Solutions,Ref1,Ref2,Ref3}.

\subsection{Solutions with one components}

The first and most simple case is when only one component ($WM$) is present.
Thus, (\ref{Fried_Int}) is reduced to%
\[
t\left(  a\right)  =\frac{1}{H_{0}}%
{\displaystyle\int} \frac{ada}{\sqrt{\Omega_{1}\left[
a^{2}+b^{2}\right]  ^{1/2}-\Omega_{\kappa 0}a^{2}}}.
\]
The solution for flat curvature ($\kappa=0$) is given by%
\begin{equation}
t_{f}\left(  a\right)  =\frac{2}{3H_{0}\sqrt{\Omega_{1}}}\left[  a^{2}%
+b^{2}\right]  ^{3/4}+\bar{t}\Rightarrow a_{f}\left(  t\right)  =\sqrt{\left[
\frac{3}{2}\sqrt{\Omega_{1}}H_{0}\left(  t-\bar{t}\right)  \right]
^{4/3}-b^{2}}\label{RRG}%
\end{equation}
where $\bar{t}$ is the integration constant and $\Omega_{1}\left(
1+b^{2}\right)  ^{1/2}=1$. This result was first derived in \cite{RRG1} and it
is presented here only for completeness.

The structure of solution for non-flat cases ($\kappa=\pm1$) is as complicated
as the solutions involving $\gamma$ and $WM$. Therefore, it will be presented
in the section \ref{two components}.

\subsection{Solutions with two components\label{two components}}

Solutions with two components are of type ($WM,\gamma$), ($WM,CM$) and
($WM,\Lambda$) with and without curvature. Because of their complicated
structure, the cases ($WM,\Lambda$) with and without curvature will be treated
in the section \ref{three components}.\ Let's perform the analysis of the two
other cases.

\subsubsection{Universe with $WM$ and $\gamma$}

For an universe constituted by $WM$ and $\gamma$, the equation
(\ref{Fried_Int}) is reduced to%
\[
t\left(  a\right)  =\pm\frac{1}{H_{0}} {\displaystyle\int}
\frac{ada}{\sqrt{\Omega_{\gamma0}+\Omega_{1}\left[
a^{2}+b^{2}\right] ^{1/2}-\Omega_{\kappa0}a^{2}}}.
\]

The solution for flat curvature is given by%
\begin{equation}
t_{f}\left(  a\right)  =\frac{1}{H_{0}}\left[  \frac{\left(  2\sqrt
{a^{2}+b^{2}}-4\Omega_{\gamma0}\right)  \sqrt{\Omega_{\gamma0}+\Omega_{1}%
\sqrt{a^{2}+b^{2}}}}{3\Omega_{1}}\right]  +\bar{t} \label{RRG_Rad_flat}%
\end{equation}
where $\bar{t}$ is the integration constant and $\Omega_{\gamma0}+\Omega
_{1}\left(  1+b^{2}\right)  ^{1/2}=1.$ Note that if we take $\Omega_{\gamma
0}=0$ the result (\ref{RRG}) is recovered. This results was first derived in
\cite{RRG1} and again it is presented here only for completeness.
Unfortunately, equation (\ref{RRG_Rad_flat}) can not be inverted and thus it
is not possible to derive an explicit solution.

For positive curvature the solution is%
\begin{equation}
t_{p}\left(  a\right)  =\frac{-1}{\left\vert \Omega_{\kappa0}\right\vert
^{3/2}H_{0}}\left[  y_{p}+\frac{\Omega_{1}}{2}Arctg\left[  \frac{\Omega
_{1}-2\left\vert \Omega_{\kappa0}\right\vert \sqrt{a^{2}+b^{2}}}{2y_{p}%
}\right]  \right]  +\bar{t} \label{RRG_Rad_Pos}%
\end{equation}
where
\[
y_{p}\equiv\sqrt{-\Omega_{\kappa0}^{2}a^{2}+\Omega_{1}\left\vert
\Omega_{\kappa0}\right\vert \sqrt{a^{2}+b^{2}}+\Omega_{\gamma0}\left\vert
\Omega_{\kappa0}\right\vert }.
\]
And for negative curvature the solution is%
\begin{equation}
t_{n}\left(  a\right)  =\frac{1}{\left\vert \Omega_{\kappa0}\right\vert
^{3/2}H_{0}}\left[  y_{n}-\frac{\Omega_{1}}{2}\ln\left[  \Omega_{1}%
+2\left\vert \Omega_{\kappa0}\right\vert \sqrt{a^{2}+b^{2}}+2y_{n}\right]
\right]  +\bar{t} \label{RRG_Rad_Neg}%
\end{equation}
where%
\[
y_{n}\equiv\sqrt{\Omega_{\kappa0}^{2}a^{2}+\Omega_{1}\left\vert \Omega
_{\kappa0}\right\vert \sqrt{a^{2}+b^{2}}+\Omega_{\gamma0}\left\vert
\Omega_{\kappa0}\right\vert }.
\]
In both cases $\bar{t}$ is an integration constant and $\Omega_{\gamma
0}+\Omega_{1}\left(  1+b^{2}\right)  ^{1/2}-\Omega_{\kappa0}=1$. As it should
be, the equations (\ref{RRG_Rad_Pos}) and (\ref{RRG_Rad_Neg}) are reduced to
(\ref{RRG_Rad_flat}) in the limit $\left\vert \Omega_{\kappa0}\right\vert
\rightarrow0$. This statement can be verified expanding the functions
$Arctg\left[  ...\right]  $ and $\ln\left[  ...\right]  $ in powers of
$\sqrt{\left\vert \Omega_{\kappa0}\right\vert }$ until third order.

\subsubsection{Universe with $WM$ and $CM$}

For a flat universe composed by $WM$ and $CM$, the equation (\ref{Fried_Int})
becomes%
\begin{equation}
t_{f}\left(  a\right)  =\pm\frac{1}{H_{0}} {\displaystyle\int}
\frac{ada}{\sqrt{\Omega_{CM0}a+\Omega_{1}\left[  a^{2}+b^{2}\right]  ^{1/2}}%
}.\label{Int_RRG_CM}%
\end{equation}
Using the definitions%
\[
y^{2}\equiv\frac{rb}{a+\sqrt{a^{2}+b^{2}}}\text{, \ }\phi\equiv Arc\sin\left(
y\right)  \text{ \ and \ }r\equiv\sqrt{\frac{\Omega_{CM0}-\Omega_{1}}%
{\Omega_{1}+\Omega_{CM0}}}\text{,}%
\]
the solution for (\ref{Int_RRG_CM}) is written as%
\begin{equation}
t_{f}\left(  a\right)  =\frac{\sqrt{2b^{3}}}{6H_{0}}\left[  \frac{\left(
\Omega_{1}+\Omega_{CM0}\right)  ^{3}}{(\Omega_{CM0}-\Omega_{1})^{5}}\right]
^{1/4}\left[  \left(  1-r^{4}\right)  F\left(  \phi,-1\right)  +\frac{\left(
r^{4}-y^{4}\right)  \sqrt{1-y^{4}}}{y^{3}}\right]  +\bar{t}\label{RRG_CM}%
\end{equation}
where $\bar{t}$ is the integration constant and $F\left(  \phi,m\right)  $ is
the elliptical integral of the first kind (see appendix \ref{Ap}). Besides, we
have the following constraint $\Omega_{CM0}+\Omega_{1}\left(  1+b^{2}\right)
^{1/2}=1.$ This solution is valid for all physical values of $\Omega_{CM0}$
and $\Omega_{1}$ avoiding $\Omega_{CM0}=\Omega_{1}$. If we take $\Omega
_{CM0}=0$ then $r=i$. Using this result and performing some simple algebra we
recover the solution (\ref{RRG}) for a universe composed only by RRG.

The structure of solution for non-flat cases is as complicated as the
solutions involving $\gamma$, $WM$ and $CM$, and thus, they will be discussed
in the next section.

\subsection{Solutions with three components\label{three components}}

Solutions with three components are of type ($WM,CM,\gamma$), ($WM,\gamma
,\Lambda$) and ($WM,CM,\Lambda$). It is not possible to achieved an analytic
solution for the case ($WM,CM,\Lambda$) because the term inside the square
root in (\ref{Fried_Int}) is a polynomial of degree greater than four.\ Let's
perform the analysis of the two other cases.

\subsubsection{Universe with $WM$, $CM$ and $\gamma$}

Suppose an universe composed by $WM$, $CM$ and $\gamma$. In this case,
equation (\ref{Fried_Int}) becomes%
\begin{equation}
t\left(  a\right)  =\pm\frac{1}{H_{0}} {\displaystyle\int}1
\frac{ada}{\sqrt{\Omega_{\gamma0}+\Omega_{CM0}a+\Omega_{1}\left[  a^{2}%
+b^{2}\right]  ^{1/2}-\Omega_{\kappa0}a^{2}}}. \label{Int_RRG_CM_Rad}%
\end{equation}
This integration can be solved for flat and non-flat cases, but for
$\kappa=\pm1$ the expressions are rather complicated. For negative and
positive curvatures the solutions involve four and seven elliptic integrals
respectively. Besides, each solution has constraint related with the
cosmological parameters. Thus, they will not be presented in this paper.

On the other hand, the flat case is relatively simpler than non-flat cases.
Indeed, if we make the following definitions%
\begin{align*}
r &  \equiv\frac{-\Omega_{\gamma0}+\sqrt{\Omega_{\gamma0}^{2}+b^{2}\left(
\Omega_{CM0}^{2}-\Omega_{1}^{2}\right)  }}{b\left(  \Omega_{1}+\Omega
_{CM0}\right)  }\text{, \ }s\equiv\frac{-\Omega_{\gamma0}-\sqrt{\Omega
_{\gamma0}^{2}+b^{2}\left(  \Omega_{CM0}^{2}-\Omega_{1}^{2}\right)  }%
}{b\left(  \Omega_{1}+\Omega_{CM0}\right)  }\\
y^{2} &  \equiv\frac{rb}{a+\sqrt{a^{2}+b^{2}}}\text{, \ \ \ \ \ }\phi\equiv
Arc\sin\left(  y\right)  \text{ \ \ \ and \ }m\equiv\frac{s}{r},
\end{align*}
the $\kappa=0$ solution of (\ref{Int_RRG_CM_Rad}) is given by%
\begin{align}
t_{f}\left(  a\right)   &  =\pm\frac{1}{6H_{0}}\sqrt{\frac{2b^{3}r^{3}%
}{\left(  \Omega_{1}+\Omega_{CM0}\right)  }}\left\{  \frac{\sqrt{\left(
y^{2}-1\right)  \left(  my^{2}-1\right)  }}{y^{3}}\left[  1+2\left(
1+m\right)  y^{2}+\frac{y^{4}}{mr^{4}}\right]  +\right.  \nonumber\\
&  \left.  +\left(  \frac{1-m^{2}r^{4}}{m^{2}r^{4}}\right)  \left[  \left(
m+2\right)  F\left(  \phi,m\right)  -2\left(  m+1\right)  E\left(
\phi,m\right)  \right]  \right\}  +\bar{t}\label{RRG_CM_Rad}%
\end{align}
where $\bar{t}$ is the integration constant and $F\left(  \phi,m\right)  $ and
$E\left(  \phi,m\right)  $ are the elliptical integral of the first and second
kind respectively (see appendix \ref{Ap}). This solution is valid for
almost\footnote{The values $b=0$ or $\Omega_{CM0}=\Omega_{1}$ are not
allowed.} all physical values of $\Omega_{\gamma0}$, $\Omega_{CM0}$,
$\Omega_{1}$ and $b$ satisfying the constraint $\Omega_{\gamma0}+\Omega
_{CM0}+\Omega_{1}\left(  1+b^{2}\right)  ^{1/2}=1$. Nevertheless, the choice
of sign depends on relation between $\Omega_{CM0}$ and $\Omega_{1}$. If
$\Omega_{CM0}>\Omega_{1}$ ($\Omega_{CM0}<\Omega_{1}$) the sign plus (minus)
must be used.

As it should be, the solution (\ref{RRG_CM_Rad}) contains the previous case
involving only $WM$ and $CM$. Taking $\Omega_{\gamma0}=0$ we get $m=-1$ and
after some straightforward algebra we recover the solution (\ref{RRG_CM}).

\subsubsection{Universe with $WM$, $\gamma$ and $\Lambda$}

For an universe constituted by $WM$, $\gamma$ and $\Lambda$ the equation
(\ref{Fried_Int}) is given by%
\begin{equation}
t\left(  a\right)  =\pm\frac{1}{H_{0}} {\displaystyle\int}
\frac{ada}{\sqrt{\Omega_{\gamma0}+\Omega_{\Lambda}a^{4}+\Omega_{1}\left[
a^{2}+b^{2}\right]  ^{1/2}-\Omega_{\kappa0}a^{2}}}. \label{Int_RRG_Rad_Lam}%
\end{equation}
It is convenient to change the variable of integration $a$ using the relation
$a^{2}=x^{2}-b^{2}$. Thus,%
\[
t\left(  a\right)  =\pm\frac{1}{H_{0}\sqrt{\Omega_{\Lambda}}}%
{\displaystyle\int}
\frac{xdx}{\sqrt{x^{4}-Lx^{2}+Mx+P}}%
\]
where%
\[
L=2b^{2}+\frac{\Omega_{\kappa0}}{\Omega_{\Lambda}},\text{ }M=\frac{\Omega_{1}%
}{\Omega_{\Lambda}}\text{ and \ }P=\frac{\Omega_{\gamma}}{\Omega_{\Lambda}%
}+b^{4}+\frac{\Omega_{\kappa0}}{\Omega_{\Lambda}}b {{}^2} .
\]

The new integral is not too simple but it can be solved through the following
steps\footnote{These steps were first developed in \cite{Solutions}.}:

\begin{enumerate}
\item Rewrite the fourth-degree polynomial in terms of the roots $r_{i}$:%
\[
x^{4}-Lx^{2}+Mx+P=\left(  x-r_{1}\right)  \left(  x-r_{2}\right)  \left(
x-r_{3}\right)  \left(  x-r_{4}\right)
\]

\item Introduce some convenient new constants:
\begin{align*}
S  &  =\sqrt[3]{2}(L^{2}+12P)\\
W  &  =-2L^{3}+27M^{2}+72LP\\
U  &  =-4(L^{2}+12P)^{3}+(-2L^{3}+27M^{2}+72LP)^{2}\\
V  &  =\left[  \frac{S}{3(W+\sqrt{U})^{1/3}}+\frac{(W+\sqrt{U})^{1/3}%
}{3\sqrt[3]{2}}\right]  +\frac{2L}{3}%
\end{align*}

\item Express the roots in terms of these constants:%
\begin{align}
r_{1}  &  =-\frac{1}{2}\left[  \sqrt{V}+\sqrt{\left(  2L-V+\frac{2M}{\sqrt{V}%
}\right)  }\right]  \text{ };\text{ }r_{2}=-\frac{1}{2}\left[  \sqrt{V}%
-\sqrt{\left(  2L-V+\frac{2M}{\sqrt{V}}\right)  }\right] \label{roots}\\
r_{3}  &  =\frac{1}{2}\left[  \sqrt{V}+\sqrt{\left(  2L-V-\frac{2M}{\sqrt{V}%
}\right)  }\right]  \text{ };\text{ }r_{4}=\frac{1}{2}\left[  \sqrt{V}%
-\sqrt{\left(  2L-V-\frac{2M}{\sqrt{V}}\right)  }\right] \nonumber
\end{align}

\item Sets the parameters $m$, $n$ and the amplitude $\phi$ as:
\begin{align*}
n  &  =\frac{r_{2}-r_{4}}{r_{1}-r_{4}}\text{, }m=n\frac{r_{1}-r_{3}}%
{r_{2}-r_{3}}\text{,}\\
\phi &  =\arcsin\left[  \sqrt{\frac{(x-r_{2})}{n(x-r_{1})}}\right]
\end{align*}

\item The solution will then be:

\begin{align}
t\left( a\right) & =\frac{2}{H_{0}}\sqrt{\frac{\left( x-r_{1}\right) ^{2}}{%
\Omega _{\Lambda }(r_{4}-r_{1})}} \left[ \frac{r_{1}\sqrt{\left(
x-r_{2}\right) }}{\sqrt{(r_{3}-r_{2})^{2}}(x-r_{1})}F\left[ \phi ,m\right]
+\right.   \notag \\
& \left. +\frac{r_{4}-r_{2}}{\sqrt{(r_{4}-r_{2})^{2}}}\sqrt{\frac{%
(r_{1}-r_{2})^{2}}{\left( r_{3}-r_{2}\right) \left( x-r_{1}\right) ^{2}}}\Pi %
\left[ n,\phi ,m\right] \right] +\bar{t}  \label{RRG_Rad_Lam}
\end{align}%
where $x=\sqrt{a^{2}+b^{2}}$, $\bar{t}$ is the integration constant and $%
F\left( \phi ,m\right) $ and $\Pi \left( n,\phi ,m\right) $ are the
elliptical integral of the first and third kind respectively (see appendix %
\ref{Ap}). Besides, we have the following constraint $\Omega _{\gamma
0}+\Omega _{\Lambda }+\Omega _{1}\left( 1+b^{2}\right) ^{1/2}-\Omega
_{\kappa 0}=1$.
\end{enumerate}

At this point, some features about this solution must be clarified. At first
sight it seems that (\ref{RRG_Rad_Lam}) could be simplified. However, as
$\left(  r_{i}-r_{j}\right)  $ and $\left(  x-r_{i}\right)  $ could be complex
numbers, any extra desirable simplification must be done with caution and only
when the values of $\Omega_{\gamma0}$, $\Omega_{\Lambda}$, $\Omega_{1}$ and
$b$ are specified. Other important point is that (\ref{RRG_Rad_Lam}) is not
valid for all physical values of $\Omega_{\gamma0}$, $\Omega_{\Lambda}$,
$\Omega_{1}$ and $b$. It happens because there is an arbitrariness in choice
of which root will be $r_{1}$, $r_{2}$, $r_{3}$ or $r_{4}$. Nevertheless, the
choice it was made in (\ref{roots}) include wide ranges for the parameters
comprising inclusively the $\Lambda CDM$ case. For practical purposes, a set
of conditions that ensure a physical solution are%
\begin{gather}
0<\Omega_{\Lambda}\leq2\text{, }0\leq\Omega_{1}\leq2\text{, }0\leq
b\leq2\label{conditions}\\
\Omega_{\gamma0}\ll\Omega_{\Lambda}\text{ \ \ and }\Omega_{\gamma0}\ll
\Omega_{1}.\nonumber
\end{gather}

\section{Final Comments}

In this paper we derived analytical solutions for a universe composed by one,
two and three components where one of them represents warm matter. The first
solution obtained is one that involving only warm matter. It is very simple
but it serves such a guide for the complex ones. The next step it was to
derive solutions containing warm matter plus radiation or cold matter. As
expected, these kind of solutions are more complicated than the previous one
and only implicit solutions were found. The most complicated solution which
were achieved are ones involving warm matter, radiation and cosmological
constant or warm matter, radiation and cold matter. These type of solutions,
with three components, always involving elliptic integrals. Unfortunately, an
analytic solution containing all the four components was not obtained.

The warm matter could mimic dark matter, neutrinos and even baryonic
matter. Thus, these solutions can be apply in different context. For
example, we can use them to analyze the effects of warm dark matter
in structure formation \cite{RRG2}. Other possibility is use them to
study massive neutrinos in cosmology. It is noteworthy that although
the results obtained concerns only to the background, they are also
important in perturbative cosmology. Indeed, the perturbative
analysis becomes simpler when the analytical solution for the
background is known.

Finally, it is important to emphasize that the warm matter is represented by
the RRG model which is an approximation for a classical relativistic gas. In
the context of thermodynamics, this approximation differs from the real
situation at most 2.5\% \cite{RRG1}. Nevertheless, none comparison was done at
cosmological context. We expect to explore this issue in the near future.

\acknowledgments

The author would like to thank FAPERN-Brazil for financial support.

\appendix

\section{Appendix\label{Ap}}

Definition of elliptic integrals \cite{Grad}:

\begin{itemize}
\item First kind:%
\[
F\left(  \phi,m\right)  = {\displaystyle\int\limits_{0}^{\sin\phi}}
\frac{dx}{\sqrt{\left(  1-x^{2}\right)  \left(  1-mx^{2}\right)  }}.
\]

\item Second kind:%
\[
E\left(  \phi,m\right)  = {\displaystyle\int\limits_{0}^{\sin\phi}}
\sqrt{\frac{\left(  1-mx^{2}\right)  }{\sqrt{\left(  1-x^{2}\right)  }}}dx.
\]

\item Third kind:%
\[
\Pi\left(  n,\phi,m\right)  = {\displaystyle\int\limits_{0}^{\sin\phi}}
\frac{dx}{\left(  1-nx^{2}\right)  \sqrt{\left(  1-x^{2}\right)  \left(
1-mx^{2}\right)  }}.
\]

\end{itemize}


\bigskip

\begin{thebibliography}{99}                                                                                               %


\bibitem {SN1}S. Perlmutter et al., Nature \textbf{391}, 51 (1998)
[astro-ph/9712212].

\bibitem {SN1A} S. Perlmutter et al., Astrophys. J. \textbf{517}, 565
(1999) [astro-ph/9812133].

\bibitem {SN2}A. G. Riess et al., Astron. J. \textbf{116}, 1009 (1998)
[astro-ph/9805201].

\bibitem {SN2A} A. G. Riess, et al., Astron. J. \textbf{118}, 2668 (1999) [astro-ph/9907038].

\bibitem {SN3}R. Amanullah et al., Astrophys. J., \textbf{716}, 712 (2010)
[arxiv:1004.1711].

\bibitem {CMB1}E. Komatsu et al., Astrophys. J. Suppl. \textbf{192}, 18 (2011)
[arxiv:1001.4538].

\bibitem {Sur1}Will J. Percival et al, Mon. Not. Roy. Astron. Soc.
\textbf{337}, 1068 (2002) [astro-ph/0206256].

\bibitem {Sur2}D. J. Eisenstein et al, Astrophys. J. \textbf{633}, 560 (2005) [astro-ph/0501171].

\bibitem {Sur3}W. J. Percival et al., Mon. Not. Roy. Astron. Soc.
\textbf{401}, 2148 (2010) [arxiv:0907.1660].

\bibitem {Solutions}R. Aldrovandi, R. R. Cuzinatto and L. G. Medeiros, Found.
Phys. \textbf{36}, 1736 (2006) [gr-qc/0508073].

\bibitem {Ref1}R. Coquereaux and A. Grossmann, Ann. Phys. \textbf{143}, 296 (1982).

\bibitem {Ref2}M. Dabrowski and J. Stelmach, Ann. Phys. \textbf{166}, 422 (1986).

\bibitem {Ref3}M. P. Dabrowski, Ann. Phys. \textbf{248}, 199 (1996).

\bibitem {RRG1}G. de Berredo-Peixoto, I. L. Shapiro and F. Sobreira, Mod.
Phys. Lett. \textbf{A} \textbf{20}, 2723 (2005) [gr-qc/0412050].

\bibitem {RRG2}J.C. Fabris, I.L. Shapiro and F. Sobreira, JCAP \textbf{02},
001 (2009) [arxiv:0806.1969].

\bibitem {RRG3}Julio C. Fabris, Ilya L. Shapiro and A. M. Velasquez-Toribio,
Phys. Rev. \textbf{D} \textbf{85}, 023506 (2012) [arxiv:1105.2275].

\bibitem {Grad}I. S. Gradshteyn and I. M. Ryzhik, \textit{Table of Integrals,
Series and Products}, 7th edn. (Academic Press, Amsterdam, 2007, edition by A.
Jeffrey and D. Zwillinger,).
\end{thebibliography}
\end{document}